\documentclass[journal, twoside]{IEEEtran}
\usepackage{indentfirst}
\usepackage{graphicx}
\usepackage{amsmath}
\usepackage{amssymb}
\usepackage{amsfonts}
\usepackage{mathrsfs}
\usepackage{leftidx}
\usepackage{color}
\usepackage{amsmath}
\usepackage{arydshln}
\usepackage{amsthm}
\usepackage{ragged2e}
\usepackage{cite}
\usepackage{enumerate}
\usepackage{longtable}
\usepackage{float}
\usepackage{stfloats}
\usepackage{hyperref}
\usepackage{algpseudocode}
\usepackage{algorithm}
\usepackage[caption=false,font=footnotesize]{subfig}
\usepackage{tabularx}
\usepackage{makecell}
\usepackage{url}
\usepackage{tabularx}

\usepackage{multirow}
\usepackage{booktabs}
\theoremstyle{plain}

\usepackage{caption}
\usepackage[table]{xcolor}
\usepackage{multirow}
\usepackage{array}

\newcolumntype{P}[1]{>{\raggedright\arraybackslash\footnotesize}m{#1}}
\newcolumntype{A}[1]{>{\centering\arraybackslash\footnotesize}m{#1}}

\usepackage[table,usenames,dvipsnames]{xcolor}
\definecolor{aa}{RGB}{175,238,238}
\definecolor{bb}{RGB}{255,255,255}

\usepackage{bm}
\usepackage{makecell}

\begin{document}

\title{Semantic-based Internet of Embodied Intelligence: Visions and Frontiers}

\author{Yaheng Wang, Rui Meng,~\IEEEmembership{Member,~IEEE,} Xiaodong Xu,~\IEEEmembership{Senior Member,~IEEE,} Yiming Liu,~\IEEEmembership{Member,~IEEE,} Feiliang Song, Linyuan Hu, Huishi Song, Lexi Xu,~\IEEEmembership{Senior Member,~IEEE,} Tony Q. S. Quek,~\IEEEmembership{Fellow,~IEEE,}

and 
Ping Zhang,~\IEEEmembership{Fellow,~IEEE}

\thanks{
This work was supported in part by the National Key R\&D Program of China under Grant 2025YFF0514404; in part by the National Natural Science Foundation of China under Grant 62501066 and under Grant U24B20131; and in part by the S\&T Program of Hebei under Grant 262X0405D.
\textit{(Corresponding author: Rui Meng and Xiaodong Xu.)}

Yaheng Wang is with the State Key Laboratory of Networking and Switching Technology, Beijing University of Posts and Telecommunications, Beijing 100876, China, and also with 
the ZGC Institute of Ubiquitous-X Innovation and Applications, Beijing 100083, China (e-mail: wangyaheng@bupt.edu.cn).

Rui Meng and Xiaodong Xu are with the State Key Laboratory of Networking and Switching Technology, Beijing University of Posts and Telecommunications, Beijing 100876, China, and also with the Satellite Internet Testing Center, Xiong'an Aerospace Information Research Institute, Xiong'an 070001, China (e-mail: buptmengrui@bupt.edu.cn; xuxiaodong@bupt.edu.cn).

Yiming Liu, Feiliang Song, Linyuan Hu, and Ping Zhang are the State Key Laboratory of Networking and Switching Technology, Beijing University of Posts and Telecommunications, Beijing 100876, China (e-mail: liuyiming@bupt.edu.cn; songfeiliang@bupt.edu.cn; hulinyuan@bupt.edu.cn; pzhang@bupt.edu.cn).

Huishi Song is with the ZGC Institute of Ubiquitous-X Innovation and Applications, Beijing 100083, China (e-mail: songhuishi@zgc-xnet.com).

Lexi Xu is with the Research Institute, China United Network Communications Corporation, Beijing 100048, China (e-mail: xulx29@chinaunicom.cn).

Tony. Q. S. Quek is with the Singapore University of Technology and Design, Singapore 487372, and also with the Department of Electronic Engineering, Kyung Hee University, Yongin 17104, South Korea (e-mail: tonyquek@sutd.edu.sg).

}}






\maketitle

\begin{abstract}
Recent advances in generative artificial intelligence (AI) and embodied intelligence (EI) enable autonomous agents to interact with the physical world. However, scaling these systems into networks of multiple agents, namely the Internet of EI (IoEI), faces critical bottlenecks. These include the overhead of massive multimodal data transmission and the decoupling of logical reasoning from physical constraints. To address these challenges, we envision the Semantic-based IoEI (SIoEI), which leverages semantic information as a unified metric throughout the agent lifecycle. We systematically define four key dimensions of EI: perception, intelligence, control, and communication. We further elaborate how semantic empowerment revolutionizes environmental perception, cognition and task planning, action generation and robust control, and communication and networking. We also present a case study to verify that, the semantic-empowered end-to-end process significantly improves channel robustness and reduces end-to-end latency for EI. Finally, we outline critical open research directions for the SIoEI paradigm.
\end{abstract}


\section{Introduction}
In recent years, breakthroughs in generative artificial intelligence (AI), such as large language models (LLMs) and generative diffusion models, have reshaped the digital landscape. However, these models excel only at processing symbols and statistical patterns in the virtual world, lacking genuine perception and interaction capabilities in the physical world. To bridge this gap, embodied intelligence (EI) integrates AI algorithms into physical entities such as robots and drones, enabling them to perceive, act, and learn, and has quickly become the next frontier in AI evolution\cite{duan2022survey}. When tasks require the collaboration of hundreds or thousands of heterogeneous robots or sensors, the need for universal connectivity catalyzes the emergence of the Internet of EI (IoEI)\cite{kountouris2021semantics}.

Nevertheless, IoEI faces two critical challenges. First, transmitting massive multimodal raw data, including images, point clouds, and audio, overloads radio access networks and hinders real-time collaboration\cite{meng2026semantic}. Second, traditional AI lacks understanding of physical properties and causal relationships, often producing decisions that violate physical laws. In response, semantic communication conveys task-relevant semantic information among embodied agents instead of raw data, drastically reducing data volume while preserving task meaning\cite{zhang2026towards, xie2021deep}. This brings the following enhancements to the four core dimensions of IoEI:
\begin{itemize}
\item \textbf{Environment Perception Dimension:} Instead of merely acquiring data, the system obtains structured semantic descriptions of scenes that support reasoning, providing input for subsequent planning.
\item \textbf{Cognition and Task Planning Dimension:} High-level goals are decomposed into subtasks via semantic rules rather than rule-based or statistical learning, making planning more physically plausible and robust to uncertainty.
\item \textbf{Action Generation and Control Dimension:} Intentional decisions can directly and flexibly influence low-level actions, while the current physical state is simultaneously fed back as semantic information.
\item \textbf{Communication and Network Coordination Dimension:} Even under bandwidth constraints and harsh channel conditions, embodied agents can maintain efficient and reliable collaboration.
\end{itemize}

Despite the disruptive potential of semantic-based IoEI (SIoEI), it remains an emerging field with major challenges. Semantic features and models trained in simulation may mismatch real physical systems. Semantic extraction or reasoning modules may generate erroneous information that propagates across the network, causing abnormal behavior. In unstructured environments, semantic decisions may lack strict physical safety boundaries, leading to collisions or falls. This paper systematically analyzes how semantics can be applied in SIoEI. Our main contributions are as follows.
\begin{itemize}
\item We propose the vision of SIoEI and define its four key dimensions, including perception, intelligence, control, and communication, that distinguish it from traditional disembodied AI.
\item We systematically analyze how semantics empowers the four phases of IoEI, from environmental perception and task planning to action generation and network coordination, and we introduce the SIoEI concept by extending EI to systematic clusters of the IoEI.
\item We present a case study and discuss future opportunities and challenges in this emerging field.
\end{itemize}

\section{Four-Dimensional Comparison Under Semantic Empowerment}

\begin{table*}
	\centering
	\caption{Four-Dimensional Comparison Between Traditional Paradigms and SIoEI}
	\label{tab:four_dimensional_comparison}
	\renewcommand{\arraystretch}{1.35}
	\begin{tabular}{|m{1.9cm}|m{4.8cm}|m{6.2cm}|m{3.6cm}|}
		\hline
		\textbf{Dimension} & \textbf{Traditional Paradigm} & \textbf{SIoEI} & \textbf{Key Implication} \\ \hline
		Perception & Passively acquires raw multimodal data, while physical attributes and task affordances remain weakly represented. & Integrates multimodal sensory inputs with proprioception and action feedback, transforming fragmented data into semantic objects with physical attributes and affordance information\cite{radford2021learning}. & From passive data acquisition to task-driven environmental understanding. \\ \hline
		Intelligence & Relies on statistical correlations in symbolic or textual spaces, lacking physical grounding and causal feedback. & Grounds reasoning in generative world models and physical common sense, enabling agents to predict environmental and network dynamics before execution\cite{fan2026generative}. & From disembodied symbolic reasoning to physically grounded behavioral reasoning. \\ \hline
		Control & Depends on accurate models and predefined trajectories, showing limited adaptability in unstructured environments. & Maps task intentions and action symbols into executable motor policies, supporting dexterous manipulation, compliant control, and self-correction\cite{ahn2022saycan}. & From trajectory tracking to intention-guided adaptive execution. \\ \hline
		Communication & Focuses on raw data transmission, causing bandwidth pressure and weak coupling with task execution. & Transmits task-relevant semantic information, action tokens, and intent representations through joint semantic coding and network coordination\cite{bourtsoulatze2019deep, dai2022nonlinear}. & From syntactic information delivery to pragmatic intent-action synergy. \\ \hline
	\end{tabular}
\end{table*}

The IoEI achieves a paradigmatic leap from information interconnection to the interconnection of intent and action. Traditional architectures face severe challenges when handling embodied tasks, including decoupled perception and action, the logic-physics barrier, and bandwidth crises caused by massive multimodal data. This section analyzes the fundamental differences between IoEI and traditional paradigms across four dimensions, as summarized in Tab. \ref{tab:four_dimensional_comparison}. The comparison shows that semantics connect heterogeneous agents and bridge the digital-physical gap. Specifically, perception shifts from passive sensing to task-driven understanding, intelligence from symbolic reasoning to physically grounded behavioral reasoning, control from trajectory tracking to intention-guided adaptive execution, and communication from syntactic delivery to pragmatic intent-action synergy.

\subsection{Perception Dimension}
Traditional sensor-based perception objectively measures the physical world and treats targets as static data sources. This passive, disembodied paradigm provides only shallow representations stripped of physical attributes such as mass or friction, and it remains decoupled from the agent's state.
In contrast, embodied perception focuses on multimodal representation and action affordance. By coupling proprioception and exteroception, agents actively change poses to verify interactive attributes. Semantic empowerment transforms fragmented multimodal data into unified, high-level semantic objects embedded with physical causality\cite{radford2021learning}, achieving task-driven intentional perception and overcoming the lack of physical common sense in traditional sensors.

\subsection{Intelligence Dimension}
Traditional LLMs derive intelligence from statistical modeling of static text and lack authentic experience of physical laws. Although they excel at language understanding and reasoning, they remain isolated from sensory feedback and physical causality.
Conversely, EI emphasizes that intelligence evolves through dynamic interaction within the perception-action loop. It utilizes multimodal foundation models for end-to-end semantic mapping of sensory inputs and employs generative diffusion models as world models\cite{fan2026generative} to predict environmental dynamics, network states, and action consequences before execution. Semantic empowerment anchors high-level logical reasoning onto physical common sense, realizing a leap from symbolic intelligence to behavioral intelligence.

\subsection{Control Dimension}
Traditional control relies on analytical geometry and classical mechanics to perform high-precision trajectory tracking in structured environments. This rigid paradigm treats actuators as pure execution mechanisms dependent on precise environmental priors and easily fails under unstructured perturbations.
Embodied control moves toward semantic policy generation and dexterous manipulation\cite{ahn2022saycan}. It couples low-level motor drives with high-level cognitive intentions, forming affordance-oriented interaction logic. Semantic empowerment maps discrete action symbols into continuous physical torques and introduces whole-body coordination and compliant control. This endows the system with physical adaptability, cross-scene generalization, and self-healing abilities.

\subsection{Communication Dimension}
Traditional communication follows Shannon's theory, pursuing lossless bit restoration. This decoupled paradigm ignores the behavioral impact of data on the physical world, causing bandwidth bottlenecks and failing to guarantee the ultra-low latency and high reliability required for embodied coordination.
In contrast, embodied communication shifts from syntactic transmission to pragmatic synergy. It is embedded in the perception-action loop and transmits task intents via action tokens rather than full raw data. Through joint source-channel semantic coding\cite{bourtsoulatze2019deep, dai2022nonlinear}, discrete action symbols become noise-resilient semantic vectors. This deep fusion dynamically allocates resources based on task urgency, enabling heterogeneous clusters to reach intent consensus and reliable collaboration under extreme channel conditions.

\section{Semantic Empowerment Across Four Dimensions of Embodied Intelligence}

\begin{figure*}[htbp]
\centering
\includegraphics[width=0.95\textwidth]{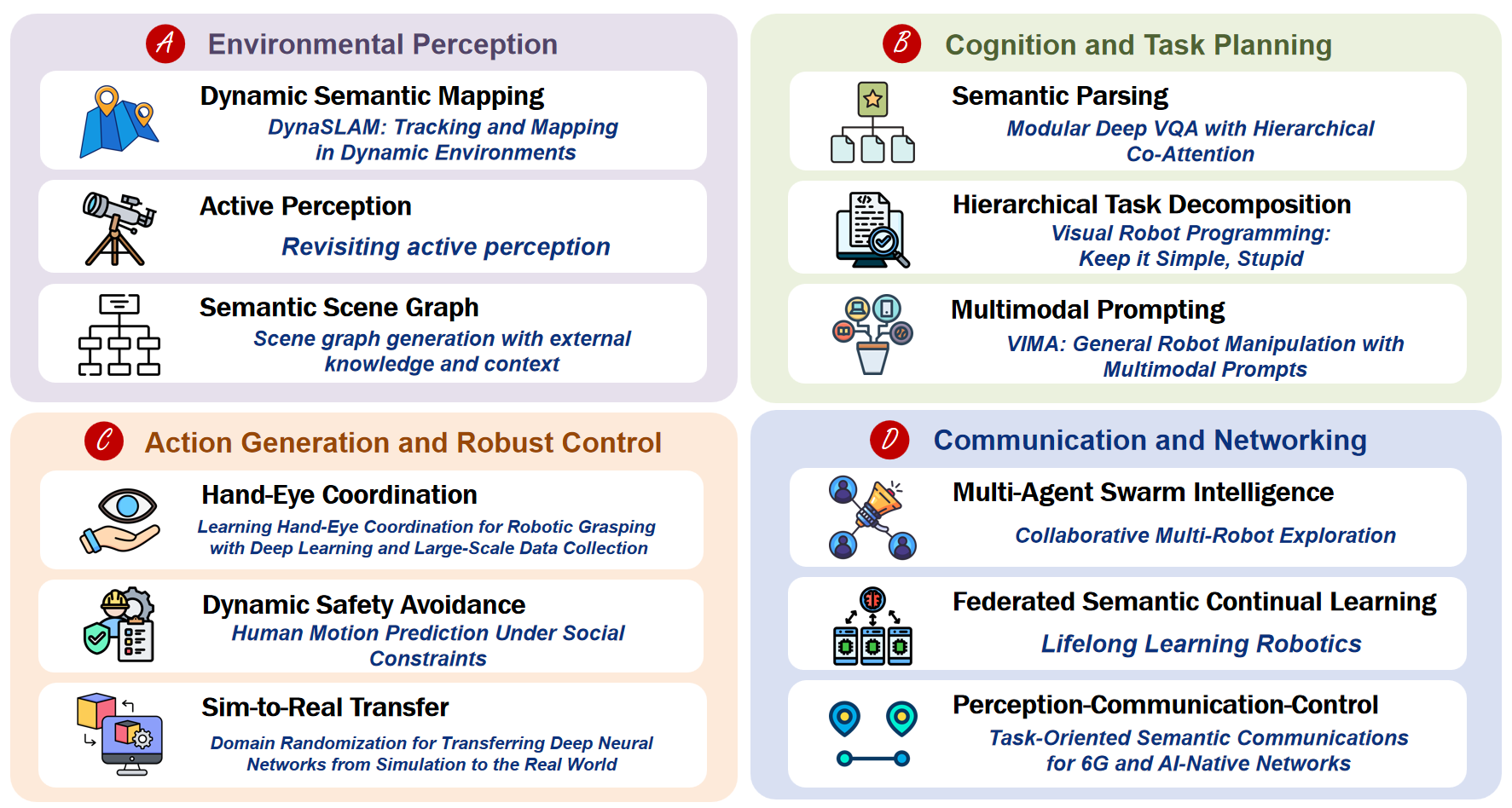}
\caption{Representative semantic-empowered technologies across the four EI dimensions: including environmental perception, cognition and task planning, action generation and robust control, and communication and networking.
}
\label{fig:four_group}
\end{figure*}

The core advantage of SIoEI lies in using semantic information as a unified metric throughout the agent-physical world interaction lifecycle. This section explores how semantics empowers perception, cognition, control, and communication. By constructing an end-to-end semantic closed loop, SIoEI reduces computational and communication redundancy while enhancing generalization and operational stability in unstructured environments, achieving a breakthrough from syntactic data processing to pragmatic task execution. 
Technical details are given in Fig. \ref{fig:four_group}.

\begin{table*}[htbp]
    \centering
    \caption{Semantic-Empowered Technologies and Methods Across Four Dimensions of SIoEI}
    \label{tab:semantic_empowerment_methods}
    \renewcommand{\arraystretch}{1.45}
    \begin{tabular}{|m{2.5cm}|m{7.0cm}|m{7.0cm}|}
        \hline
        \textbf{Dimension} & \textbf{Technologies} & \textbf{Methods} \\ \hline

        \multirow{3}{=}{\raggedright Environmental Perception}
        & Multi-modal Perception and Data Fusion
        & Align visual, LiDAR, tactile, and language features into a shared semantic space with task-driven sensor attention \cite{radford2021learning}. \\ \cline{2-3}

        & 3D Semantic Mapping and Localization
        & Build semantic maps and scene graphs; identify affordances and open-vocabulary objects \cite{rosinol2021kimera, gu2022open}. \\ \cline{2-3}

        & Fine-grained Object Recognition and Attribute Understanding
        & Identify unseen objects via language models; extract spatial relationships and physical-state monitoring semantics \cite{gu2022open}. \\ \hline

        \multirow{4}{=}{\raggedright Cognition and Task Planning}
        & Semantic Parsing of Natural Language Instructions
        & Parse vague instructions into executable primitives and infer implicit steps \cite{ahn2022saycan}. \\ \cline{2-3}

        & Hierarchical Semantic Decomposition of Long-horizon Tasks
        & Decompose large-scale goals into standardized action units and verify logical dependencies \cite{ahn2022saycan}. \\ \cline{2-3}

        & Knowledge Base Retrieval via Commonsense Reasoning
        & Use commonsense knowledge for conflict detection and cross-domain knowledge transfer \cite{driess2023palme}. \\ \cline{2-3}

        & Generative Diffusion Models for Environmental and Network Prediction
        & Predict channel states, traffic patterns, and environmental evolution to support robust planning and semantic transmission \cite{fan2026generative}. \\ \hline

        \multirow{3}{=}{\raggedright Action Generation and Robust Control}
        & Agentic AI for Action Skill Generation and Precision Manipulation
        & Decompose high-level intentions into executable actions and adapt manipulation policies through feedback-driven agentification \cite{zhang2026toward}. \\ \cline{2-3}

        & Semantic-oriented Dynamic Safety Avoidance
        & Set dynamic safety thresholds, predict interaction intent, and perform compliant correction. \\ \cline{2-3}

        & Sim-to-Real Transfer
        & Extract invariant semantic features; map sensor inputs to control tokens via end-to-end semantic modeling. \\ \hline

        \multirow{3}{=}{\raggedright Communication and Networking Coordination}
        & Multi-Agent Swarm Intelligence Collaboration
        & Align task semantics among agents and dispatch semantic instructions rather than raw data \cite{kountouris2021semantics}. \\ \cline{2-3}

        & Cross-Device Knowledge Sharing and Continual Learning
        & Build semantic experience pools and exchange model parameters under privacy constraints. \\ \cline{2-3}

        & Semantic-based Resource Scheduling
        & Allocate resources by semantic importance and channel conditions; compress content semantics on demand \cite{bourtsoulatze2019deep, dai2022nonlinear}. \\ \hline
    \end{tabular}
\end{table*}

\subsection{Environmental Perception}
Environmental perception is the starting point for embodied agents to interact with the physical world, transforming fragmented sensor data into structured semantic information.

\subsubsection{Multi-modal Perception and Data Fusion}
Semantic empowerment enables deep integration of full-sensory environmental representations. Cross-modal semantic alignment associates visual features, LiDAR point clouds, and tactile feedback within a unified latent semantic space\cite{radford2021learning}, enabling understanding of physical attributes beyond geometry. The perception system can then label affordances during perception to directly guide subsequent actions. A semantic-driven attention mechanism dynamically adjusts sensor sampling rates and weights based on task priority, reducing data processing pressure while ensuring critical information acquisition.

\subsubsection{3D Semantic Mapping and Localization}
By integrating semantic labels into Simultaneous Localization and Mapping, SIoEI constructs dynamic semantic maps with logical topological relationships\cite{rosinol2021kimera}. Topological semantic modeling partitions physical space into functionally defined areas, enhancing global path planning efficiency. Stable semantic landmarks assist localization and mitigate long-term drift. Dynamic object semantic filtering identifies and predicts the intentions of pedestrians or vehicles, removing temporary occlusions during map construction to ensure map purity and reliability.

\subsubsection{Fine-grained Object Recognition and Attribute Understanding}
Semantics empowers perception to transcend simple category recognition and delve into physical attributes and interactive relationships. Open-vocabulary semantic recognition\cite{gu2022open} allows agents to identify unseen objects by associating unknown visual features with known concepts via language models. Spatial relationship semanticization extracts relative positional semantics to support complex logic tasks. Physical state monitoring semantics perceive real-time dynamic changes as task-triggering signals, enabling a leap from static observation to state-closed-loop interaction.

\subsection{Cognition and Task Planning}

Cognition and task planning is the central control unit of EI, responsible for high-level logical reasoning and transforming vague human intentions into executable, logically complete task blueprints.

\subsubsection{Semantic Parsing of Natural Language Instructions}
Semantic empowerment maps unstructured instructions to logical primitives. An intent completion mechanism infers implicit steps for underspecified commands, while contextual disambiguation uses environmental perception to resolve pronoun references or polysemous terms.

\subsubsection{Hierarchical Semantic Decomposition of Long-horizon Tasks}
For complex long-horizon tasks, semantic empowerment ensures continuity of action sequences. A semantic primitive library decomposes large-scale goals into standardized action units, enhancing plan reusability. The SayCan scheme proposed in \cite{ahn2022saycan} synergizes LLM semantic planning with affordance-based execution probabilities of low-level skills, ensuring logical coherence. Logical dependency verification checks temporal constraints between steps, and task-progress semantic anchoring supports precise recovery after interruptions or unexpected environmental changes.

\subsubsection{Knowledge Base Retrieval via Commonsense Reasoning}
Integrating large-scale external semantic knowledge allows agents to possess human-like commonsense judgment. Intuitive semantic heuristics use prior knowledge to narrow search spaces and improve efficiency. The PaLM-E model proposed in \cite{driess2023palme} embeds multimodal perception data directly into the language model latent space, enabling cross-domain knowledge transfer and leveraging internet-scale data for novel physical tasks. Dynamic semantic conflict detection identifies states that violate physical common sense in real time and triggers immediate re-planning.

\subsubsection{World Models and Environmental Evolution Prediction}
Semantic empowerment gives agents foresight to anticipate action consequences. Generative diffusion models\cite{fan2026generative} can synthesize network states, channel evolution, and environmental dynamics, enabling agents to evaluate candidate plans before physical execution. This predictive capability supports robust resource orchestration and semantic transmission in SIoEI, and it provides a foundation for multi-agent collaboration.

\subsection{Action Generation and Robust Control}

Action generation and robust control transforms thought into action, focusing on precision, safety, and adaptability in complex physical environments. Through semantic empowerment, the control process advances from numerical computation to logic-driven execution.

\subsubsection{Action Skill Generation and Precision Manipulation}
Semantic empowerment endows low-level actions with physical common sense, moving from pure geometric modeling to intent-driven action selection. Agentic AI\cite{zhang2026toward} enables agents to decompose high-level intentions into executable manipulation actions, select appropriate tools or policies, and refine motor skills through environmental feedback. This improves manipulation success for heterogeneous objects in unstructured environments. Action symbolic control encapsulates instructions into semantic symbols such as place gently, and tactile semantic closed-loop feedback enables precision adjustment and dynamic force correction.

\subsubsection{Semantic-oriented Dynamic Safety Avoidance}
During execution, semantic information provides a hierarchical reference for safety assurance. Target-category semantic avoidance sets dynamic safety thresholds based on obstacle levels, allowing robots to adjust avoidance margins for different risk sources. Interaction intent prediction captures pedestrian-robot interactive semantics through attention mechanisms, enabling precise pre-judgment and proactive avoidance. Path semantic compliance checks ensure trajectories follow situational rules, while semanticized anomaly correction identifies execution deviations and triggers real-time compliant control compensation, enabling intelligent self-healing.

\subsubsection{Sim-to-Real Transfer}
As a unified metric, semantics bridges the Sim-to-Real transfer gap. Semantic feature invariance extracts shared features across virtual and real domains, reducing the impact of lighting or material differences on policy transfer. It maps sensor inputs directly to control semantic tokens through end-to-end semantic modeling, bypassing cumulative latency from multiple format conversions in layered architectures. This fusion enables real-time semantic comparison between execution status and expected goals, reduces invalid communication, supports semantic fault retries, and ensures robust physical operations.

\subsection{Communication and Networking Coordination}

Communication and networking coordination connects isolated embodied agents through the network to achieve resource complementarity, knowledge sharing, and task synchronization, serving as the key for SIoEI to advance toward large-scale socialized operations.

\subsubsection{Multi-Agent Swarm Intelligence Collaboration}
Semantic empowerment addresses coordination challenges caused by device heterogeneity. A semantic consensus alignment mechanism ensures that all agents maintain consistent semantic understanding of the same task goal\cite{kountouris2021semantics}. Distributed semantic decision distribution dispatches high-level semantic instructions rather than raw data, greatly reducing bandwidth requirements. A lightweight distributed semantic communication framework jointly designs feature extraction and receiver networks to meet IoT power and latency constraints, enhancing task-oriented information delivery. Heterogeneous collaborative semantic protocols enable different robot types to achieve seamless synergy through unified interfaces.

\subsubsection{Cross-Device Knowledge Sharing and Continual Learning}
Through cloud networks, agents can accumulate experience globally. Semantic experience pools transform learned skills into searchable semantic libraries. A federated semantic learning framework enables skill evolution by exchanging model parameters while protecting privacy. An information-bottleneck-driven variant establishes an optimal trade-off between semantic compression rates and task execution accuracy, achieving efficient distributed semantic model collaboration with extremely low communication overhead. Incremental semantic updates synchronize only newly emerging semantic objects, realizing on-demand dynamic knowledge acquisition.

\subsubsection{Semantic-based Resource Scheduling}
Semantic empowerment achieves precise matching between communication resources and task requirements. Semantic slicing automatically allocates network priorities based on business urgency. Advanced physical-layer receiver technologies can also reshape semantic resource scheduling. Onboard Rydberg atomic quantum receivers, for example, have been shown to enhance ground-satellite direct access and sensitivity\cite{11417150}, offering a reliable foundation for semantic transmission over challenging space-air-ground links and supporting dynamic resource scheduling. Combined with semantic-aware routing, the system selects optimal paths based on content semantic importance. When bandwidth is limited, on-demand semantic compression transmits only critical semantic features, reducing unnecessary raw pixel transmissions and ensuring low latency and high reliability for critical task semantics. 

\section{Case Study: Semantic Communication and Control for Robotic Manipulation}
\label{section6}

\subsection{Motivations}
Current research on EI often assumes idealized perception and overlooks the communication bottleneck between sensors and controllers. When raw visual observations travel over bandwidth-limited wireless channels, conventional bit-level compression and channel coding introduce severe latency and reconstruction failures at low SNR, degrading closed-loop control. 

\subsection{The presented scheme}
In SIoEI systems that tightly couple perception, communication, and cognition, communication is part of the control loop rather than mere data transport. As illustrated in Fig.~\ref{fig:compare}, the presented Joint Semantic Cognition-Communication-Control (JSCCC) scheme closes this gap by embedding semantic extraction, transmission, and decision making into a single end-to-end pipeline. 
Specifically, it converts observations into compact semantic representations through a semantic extractor. A semantic codec transports them under channel constraints, and a semantic controller maps the decoded semantics directly to actuator commands, eliminating separate reconstruction and object-detection stages.

\subsection{Simulation Results and Analysis}
\subsubsection{Comparative Schemes}
\begin{itemize}
\item The baseline scheme uses JPEG compression followed by LDPC channel coding. The received image is reconstructed and fed into a hand-crafted vision-based geometric reaching (VGR) controller that outputs joint commands for the UR5 arm. 
\item The SemComm scheme processes the same image input with a SwinJSCC semantic codec and transmits task-relevant semantic features over the physical channel. The receiver decodes the semantic vector and sends it to the same VGR controller, bypassing conventional image reconstruction. 
\end{itemize}

\subsubsection{Simulation Parameters}
\begin{figure*}
\centering
\includegraphics[width=0.95\textwidth]{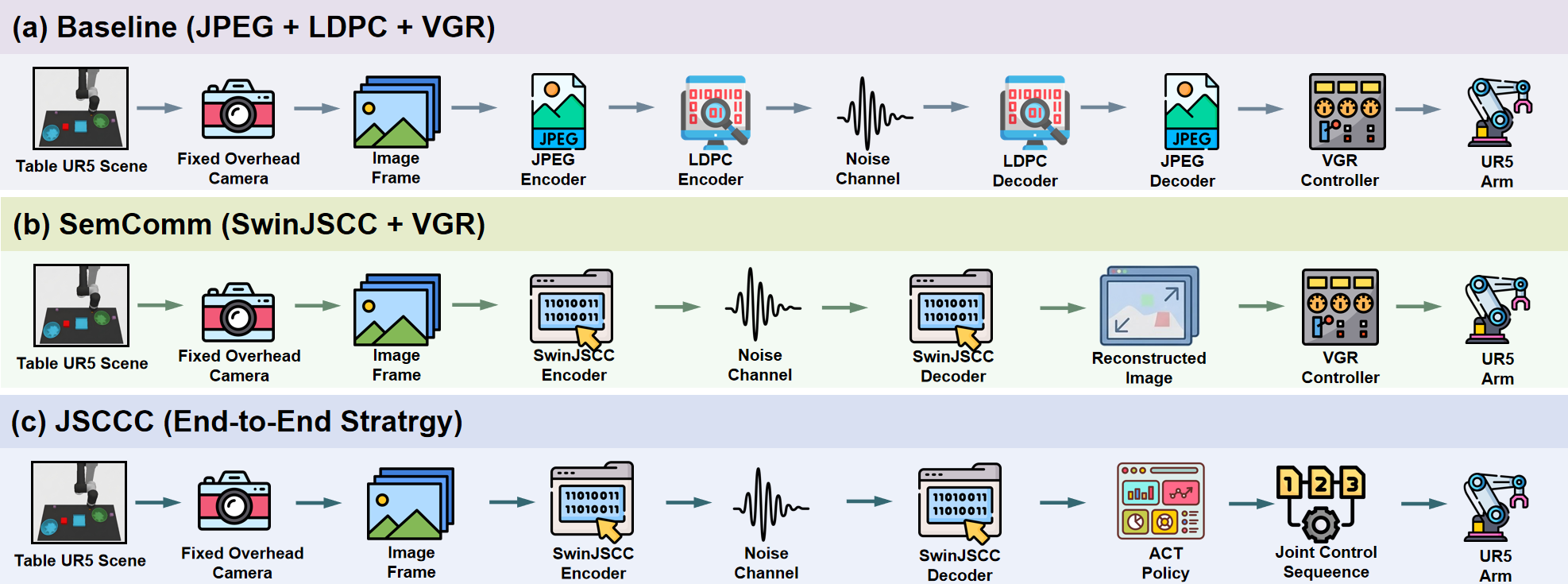}
\caption{Block diagrams of the three embodied-agent communication pipelines. (a) Baseline (JPEG+LDPC+VGR). (b) SemComm (SwinJSCC+VGR). (c) JSCCC.}
\label{fig:compare}
\end{figure*}

As shown in Fig.~\ref{fig:compare}, a 6-DoF UR5 arm grasps a colored block on a tabletop from the view of a fixed $512\times512$ RGB camera. A trial is successful when the end-effector stops within 5 cm of the target center. We compare the three schemes at SNR $=$ 5, 10, 20, 30 dB, with 47 random seeds for each.

\subsection{Results and Analysis}

\subsubsection{Task Success Rate}
\begin{figure}
\centering
\includegraphics[width=\columnwidth]{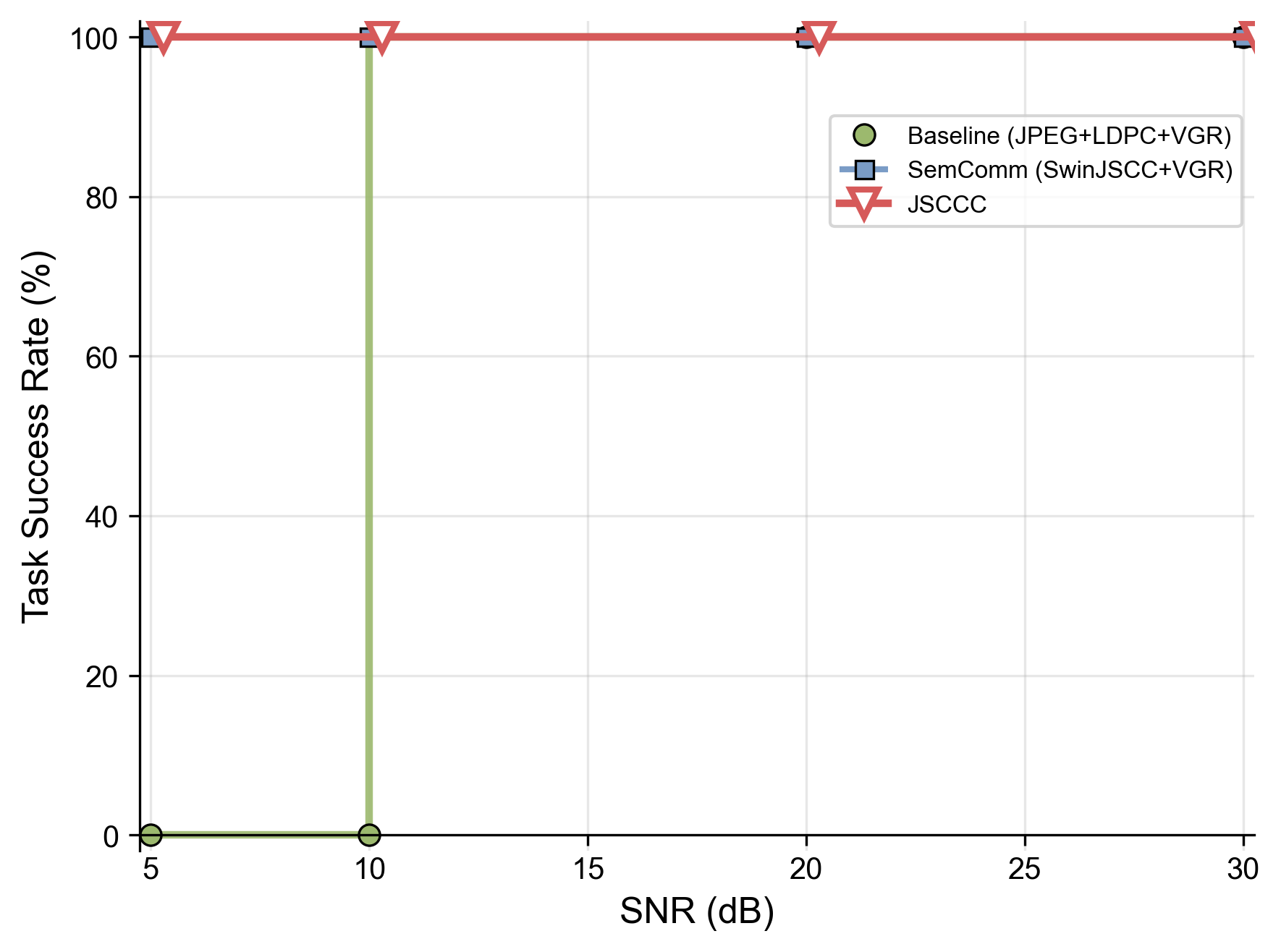}
\caption{Task success rate of the three schemes across SNR $=$ 5, 10, 20, 30 dB.}
\label{fig:success_rate}
\end{figure}

We measured the task success rate of the three schemes under SNR $=$ 5, 10, 20, 30 dB, as shown in Fig.~\ref{fig:success_rate}. The JSCCC framework achieved 100\% success at all tested SNRs because it transmits compact, task-relevant semantic features. The SemComm scheme also maintained high success, whereas the Baseline succeeded only at high SNR and collapsed to 0\% at 10 dB and below. This cliff-edge behavior reflects the fragility of bit-level reconstruction.

\subsubsection{End-to-End Latency}
\begin{figure}
\centering
\includegraphics[width=\columnwidth]{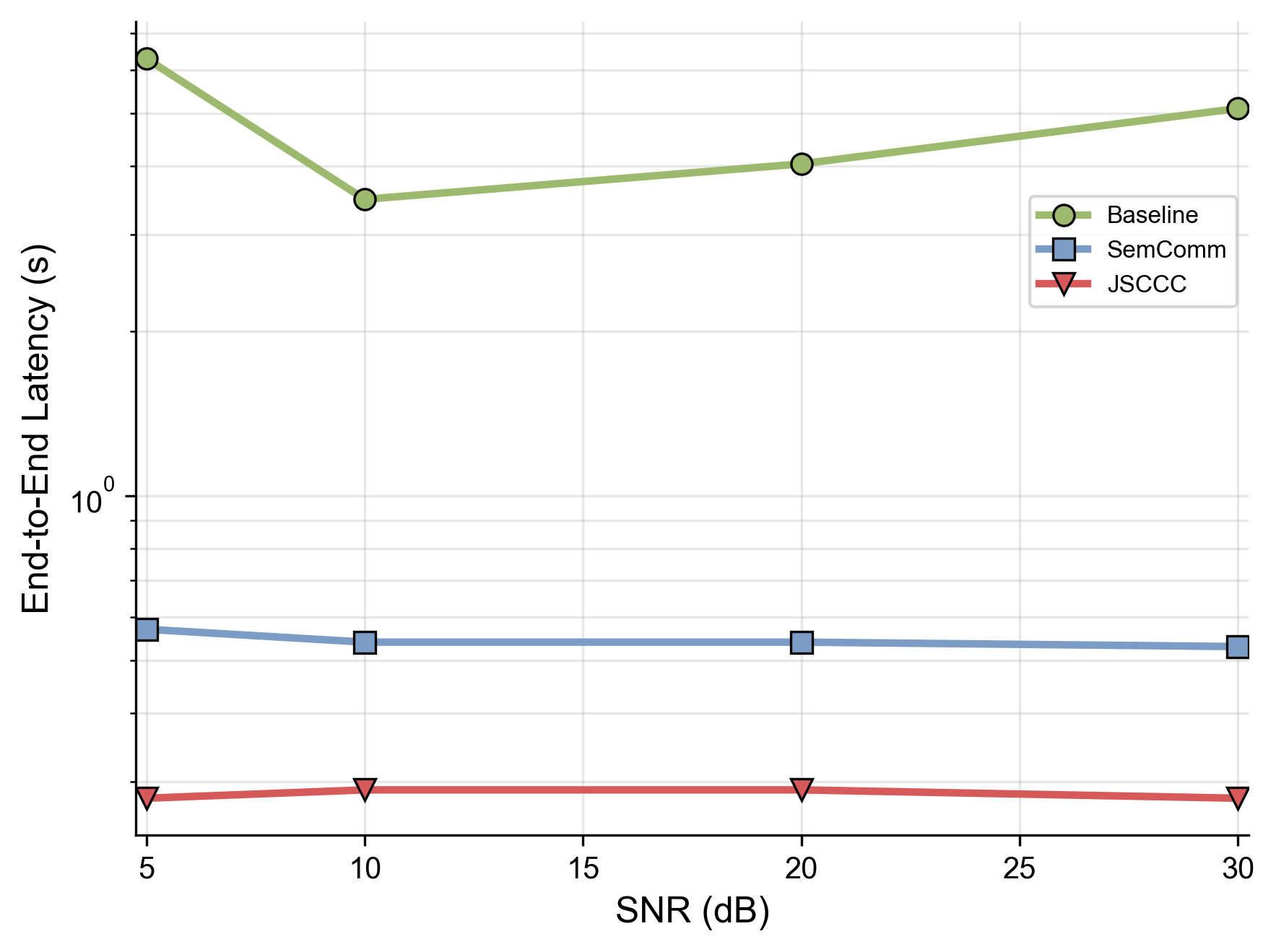}
\caption{End-to-end latency of the three schemes across SNR $=$ 5, 10, 20, 30 dB.}
\label{fig:latency}
\end{figure}

We also measured the end-to-end latency of the three schemes under the same SNR values, as shown in Fig.~\ref{fig:latency}. JSCCC and SemComm substantially reduced latency relative to the Baseline because semantic transmission avoids heavy channel coding and pixel-level reconstruction. These results show that semantic-empowered communication improves both robustness and timeliness in closed-loop robotic control.

\subsection{Lessons Learned}

The robotic manipulation case study reveals both the strengths and the limits of JSCCC under bandwidth-limited and low-SNR conditions. On the strength side, JSCCC maintained high success across all tested SNRs and significantly reduced end-to-end latency. This gain arises because semantic features are naturally robust to channel degradation, and the tight coupling of perception, communication, and control avoids pixel-level redundant transmission with low control utility. By contrast, the Baseline scheme, which relies on bit-level reconstruction, collapsed rapidly below 10 dB. This gap indicates that traditional decoupled communication-control architectures cannot meet the closed-loop reliability required by embodied intelligence.

On the limitation side, the experiments remain in simulation. The task involves a single arm reaching for a simple block, the camera pose is fixed, and the semantic extractor is assumed to be accurate. These idealized conditions differ from real-world scenarios. In practice, semantic extraction errors, dynamic occlusions, multi-agent coordination, and more complex manipulation tasks would all narrow the effective operating range of JSCCC. Moreover, the experiments do not systematically evaluate safety constraints, fault recovery, or energy trade-offs. Extending JSCCC to real robot platforms and establishing quantifiable safety boundaries are therefore important directions for future work.

\section{Conclusion}

This paper proposes the SIoEI, which treats semantic information as a unified metric throughout the lifecycle of perception, intelligence, control, and communication. Through four-dimensional comparison, we explain how semantic empowerment advances EI from passive data acquisition, disembodied symbolic reasoning, rigid trajectory tracking, and syntactic information transmission toward task-driven environmental understanding, physically grounded behavioral reasoning, intention-guided adaptive execution, and pragmatic intent-action synergy. By unifying semantics across the agent lifecycle, SIoEI offers a principled path toward scalable, resilient, and intent-aware multi-agent embodied systems. The case study further shows that the JSCCC framework maintains 100\% success at low SNR while substantially reducing end-to-end latency, whereas traditional bit-level transmission collapses quickly as the channel degrades. 

Nevertheless, SIoEI remains at an early stage. Domain gaps between simulation and real systems may cause semantic features and policies to fail during transfer, so more robust Sim-to-Real mechanisms are needed. Errors from semantic extraction and reasoning modules may amplify across the network, calling for verifiable safety boundaries and fault-tolerant mechanisms. Semantic consensus protocols, dynamic resource scheduling, and privacy-preserving knowledge sharing in multi-agent scenarios still require deeper investigation. Finally, standardized semantic ontologies, benchmarks, and large-scale experimental platforms will be essential for moving SIoEI from theoretical vision to practical deployment.

\bibliographystyle{IEEEtran}

\bibliography{ref}

@article{kountouris2021semantics,
  title={Semantics-empowered communication for networked intelligent systems},
  author={Kountouris, Marios and Pappas, Nikolaos},
  journal={IEEE Communications Magazine},
  volume={59},
  number={6},
  pages={96--102},
  year={2021},
  publisher={IEEE}
}

@article{meng2026semantic,
  title={Semantic Radio Access Networks: Architecture, State-of-the-Art, and Future Directions},
  author={Meng, Rui and Huang, Zixuan and Yan, Jingshu and Sun, Mengying and Liu, Yiming and Feng, Chenyuan and Xu, Xiaodong and Zhang, Zhidi and Gao, Song and Zhang, Ping and others},
  journal={IEEE Transactions on Cognitive Communications and Networking},
  volume={12},
  pages={7076--7097},
  year={2026},
  publisher={IEEE}
}

@article{zhang2026towards,
  title={Towards semantic-based agent communication networks: Vision, technologies, and challenges},
  author={Zhang, Ping and Meng, Rui and Xu, Xiaodong and Wang, Yaheng and Huang, Zixuan and Liu, Yiming and Zhang, Ruichen and Liu, Yinqiu and Tong, Haonan and Song, Huishi and others},
  journal={arXiv preprint arXiv:2603.24328},
  year={2026}
}

@article{xie2021deep,
  title={Deep learning enabled semantic communication systems},
  author={Xie, Huiqiang and Qin, Zhijin and Li, Geoffrey Ye and Juang, Biing-Hwang},
  journal={IEEE transactions on signal processing},
  volume={69},
  pages={2663--2675},
  year={2021},
  publisher={IEEE}
}

@article{duan2022survey,
  title={A survey of embodied ai: From simulators to research tasks},
  author={Duan, Jiafei and Yu, Samson and Tan, Hui Li and Zhu, Hongyuan and Tan, Cheston},
  journal={IEEE Transactions on Emerging Topics in Computational Intelligence},
  volume={6},
  number={2},
  pages={230--244},
  year={2022},
  publisher={IEEE}
}

@inproceedings{radford2021learning,
  author    = {Radford, Alec and Kim, Jong Wook and Hallacy, Chris and Ramesh, Aditya and Goh, Gabriel and Agarwal, Sandhini and Sastry, Girish and Askell, Amanda and Mishkin, Pamela and Clark, Jack and others},
  title     = {Learning Transferable Visual Models From Natural Language Supervision},
  booktitle = {Proceedings of the 38th International Conference on Machine Learning (ICML)},
  pages     = {8748--8763},
  year      = {2021}
}

@inproceedings{ahn2022saycan,
  author    = {Ahn, Michael and Brohan, Anthony and Brown, Noah and Chebotar, Yevgen and Cortes, Omar and David, Byron and Finn, Chelsea and Fu, Chuyuan and Gopalakrishnan, Keerthana and Hausman, Karol and others},
  title     = {Do As I Can, Not As I Say: Grounding Language in Robotic Affordances},
  booktitle = {Proceedings of the 11th International Conference on Learning Representations (ICLR)},
  year      = {2023}
}

@article{bourtsoulatze2019deep,
  title={Deep joint source-channel coding for wireless image transmission},
  author={Bourtsoulatze, Eirina and Kurka, David Burth and G{\"u}nd{\"u}z, Deniz},
  journal={IEEE Transactions on Cognitive Communications and Networking},
  volume={5},
  number={3},
  pages={567--579},
  year={2019},
  publisher={IEEE}
}

@article{dai2022nonlinear,
  author  = {Dai, Jincheng and Wang, Sixian and Tan, Kailun and Si, Zhonghua and Liu, Xiaomeng and Li, Kai and Ping, Zhang},
  title   = {Nonlinear Transform Source-Channel Coding for Semantic Communications},
  journal = {IEEE Journal on Selected Areas in Communications},
  volume  = {40},
  number  = {8},
  pages   = {2300--2316},
  year    = {2022},
  month   = aug
}

@article{rosinol2021kimera,
  author  = {Rosinol, Antoni and Abate, Marcus and Chang, Yun and Carlone, Luca},
  title   = {Kimera: From {SLAM} to Spatial Perception with {3D} Dynamic Scene Graphs},
  journal = {The International Journal of Robotics Research},
  volume  = {40},
  number  = {12--14},
  pages   = {1510--1546},
  year    = {2021}
}

@inproceedings{gu2022open,
  author    = {Gu, Xiuye and Lin, Tsung-Yi and Kuo, Weicheng and Cui, Yin},
  title     = {Open-Vocabulary Object Detection via Vision and Language Knowledge Distillation},
  booktitle = {Proceedings of the 10th International Conference on Learning Representations (ICLR)},
  year      = {2022}
}

@inproceedings{driess2023palme,
  author    = {Driess, Danny and Xia, Fei and Sajjadi, Mehdi S. M. and Lynch, Corey and Chowdhery, Aakanksha and Ichter, Brian and Wahid, Ayzaan and Tompson, Jonathan and Vuong, Quan and Yu, Tianhe and others},
  title     = {{PaLM-E}: An Embodied Multimodal Language Model},
  booktitle = {Proceedings of the 40th International Conference on Machine Learning (ICML)},
  pages     = {8469--8488},
  year      = {2023}
}

@article{fan2026generative,
  title={Generative diffusion models for wireless networks: Fundamental, architecture, and state-of-the-art},
  author={Fan, Dayu and Meng, Rui and Xu, Xiaodong and Liu, Yiming and Nan, Guoshun and Feng, Chenyuan and Han, Shujun and Gao, Song and Xu, Bingxuan and Niyato, Dusit and others},
  journal={IEEE Communications Surveys \& Tutorials},
  volume={28},
  pages={5632--5677},
  year={2026},
  publisher={IEEE}
}

@ARTICLE{11417150,
  author={Peng, Qihao and Gong, Tierui and Song, Zihang and Luo, Qu and Lin, Zihuai and Xiao, Pei and Yuen, Chau},
  journal={IEEE Wireless Communications},
  title={Enhanced Ground--Satellite Direct Access via Onboard Rydberg Atomic Quantum Receivers},
  year={2026},
  volume={33},
  number={3},
  pages={23-30},
  doi={10.1109/MWC.2026.3659837}
}

@article{zhang2026toward,
  title={Toward edge general intelligence with agentic AI and agentification: Concepts, technologies, and future directions},
  author={Zhang, Ruichen and Liu, Guangyuan and Liu, Yinqiu and Zhao, Changyuan and Wang, Jiacheng and Xu, Yunting and Niyato, Dusit and Kang, Jiawen and Li, Yonghui and Mao, Shiwen and others},
  journal={IEEE Communications Surveys \& Tutorials},
  volume={28},
  pages={4285--4318},
  year={2026},
  publisher={IEEE}
}

\end{document}